\documentclass[12pt,fleqn]{article}
\usepackage{amssymb,graphicx,epsfig}

\newcommand{\goes}{\rightarrow} 
 
\newcommand{\GeV}{\; \mathrm{GeV}} 
\newcommand{\TeV}{\; \mathrm{TeV}} 
\newcommand{\beq}{\begin{equation}} 
\newcommand{\eeq}{\end{equation}} 
\newcommand{\bea}{\begin{eqnarray}} 
\newcommand{\eea}{\end{eqnarray}}

\newcommand{\strahl}{e^+e^- \goes Z h^0 \goes \bar{\nu} \nu \, h^0}
\newcommand{\wwfus}{e^+e^- \goes \bar{\nu}_e \nu_e \,  W W 
                                  \goes \bar{\nu}_e \nu_e \,  h^0}

\newcommand{\higpro}{e^+e^- \goes \bar{\nu} \nu h^0}

\begin{document} 
%==============================================================================
%  Title
%==============================================================================
\begin{flushright}
%% \today \\
HEPHY-PUB 765/02\\  
hep-ph/0210330 
\end{flushright} 
\begin{center}

{\Large\bf\boldmath
    Radiative corrections to single Higgs boson\\ 
    production in \boldmath{$e^+e^-$} annihilation}\\[5mm]

H.~Eberl\,\footnote{Speaker}, W.~Majerotto, and V.~C.~Spanos\\[5mm]

{\it Contribution to SUSY02, 
10th International Conference on Supersymmetry 
and Unification of Fundamental Interactions, 
17--23 June 2002, DESY Hamburg, Germany.}

\end{center}

\begin{abstract}
For energies relevant to future
linear colliders, $\sqrt{s} \gtrsim 500 \GeV$, the $WW$ fusion
channel dominates the Higgs boson production 
cross section $e^+ e^- \goes \bar{\nu} \nu h^0$. 
We have calculated the one-loop corrections to this
process due to fermion and sfermion loops in the context of the MSSM. 
As a special case, 
the contribution of the fermion loops in the SM
has also been studied. 
In general, the correction is negative and sizeable of the order of $10\%$,
the bulk of it being due to fermion loops. 
\end{abstract} 

\baselineskip=17.8pt
%%%%%%%%%%%%%%%%%%%%%%%%%%%%%% Paper body %%%%%%%%%%%%%%%%%%%%%%%%
%%\section{Introduction}

No Higgs boson could be detected so far. The four LEP experiments delivered 
lower bounds~\cite{higgs} for the Standard Model (SM) Higgs mass, $m_h \gtrsim 114 \GeV$
and the light CP even Higgs boson mass $m_{h^0} \gtrsim 88.3 \GeV$ of the
Minimal Supersymmetric Standard Model (MSSM). 
In $e^+e^-$ collisions,
for energies $\gtrsim 200 \GeV$, the production of a
single Higgs boson plus missing energy starts to be dominated by
$WW$ fusion \cite{fusion,alta,kilian}, 
that is $\wwfus$, whereas the Higgsstrahlung process \cite{ellis}
$\strahl$ becomes less important.
The rates for the $ZZ$ fusion are generally one order of
magnitude smaller than those of the $WW$ channel.    

At LHC, in $p\,p$ collisions, the gluon-gluon fusion mechanism provides the
dominant contribution to Higgs production. Recently,
it has been argued that also the channels 
$WW \to h^0/H^0 \goes \tau\bar{\tau}$ 
and $WW$, can serve as suitable
search channels at LHC 
even for a Higgs boson mass of $m_h \sim 120\GeV$  \cite{zepe,mazzucato}.
It was also shown \cite{aachen} that the $e\, p$-option at LHC 
would offer the best opportunity to search for a Higgs
boson in the mass range $m_h < 140 \GeV$ with $WW$ 
(and $ZZ$) fusion as the most
important Higgs boson production mechanism there. At Tevatron, 
with $p \, \bar{p}$ collisions at $2 \TeV$, 
the $WW$ fusion process plays a less important r\^ole at least for
$m_h \lesssim 180 \GeV$ \cite{tevatron}. 

\vspace{2mm}
This contribution follows~Ref.~\cite{eberl} where
the leading one-loop corrections
to the $WWh^0$ vertex in the MSSM were calculated.
Because of their Yukawa couplings, the fermion/sfer\-mion
loops are taken into account. Then the 
total cross section in 
$e^+e^-$ annihilation was worked out for $\wwfus$.
We also included the Higgsstrahlung process $\strahl$ and the
interference between these two mechanisms.
Because the Higgsstrahlung process is much smaller in this range,
we have neglected its radiative corrections.
We have also discussed the SM case. 

%%%\section{$W$-boson fusion}
As for energies $\sqrt{s} > 500 \GeV$ the dominant channel $\higpro$
is by far the $WW$ fusion, taking into account only (s)fermion loops, 
the renormalization of the five-point function
simplifies to the renormalization of the $W W h^0$  vertex with 
off-shell $W$ bosons.
The renormalization of the other two vertices in 
the process (e.~g.  the $e^- \nu_e W^+$ coupling) is absorbed.
For the renormalization procedure the on-shell scheme has been
adopted. 

%%%%%%%%%%%%%%%%%%% Fig 1 %%%%%%%%%%%%%%%%%%%%%%%%%%%%%%%%%%%%%%%
\begin{figure}[th]  
\centering\includegraphics[scale=0.85]{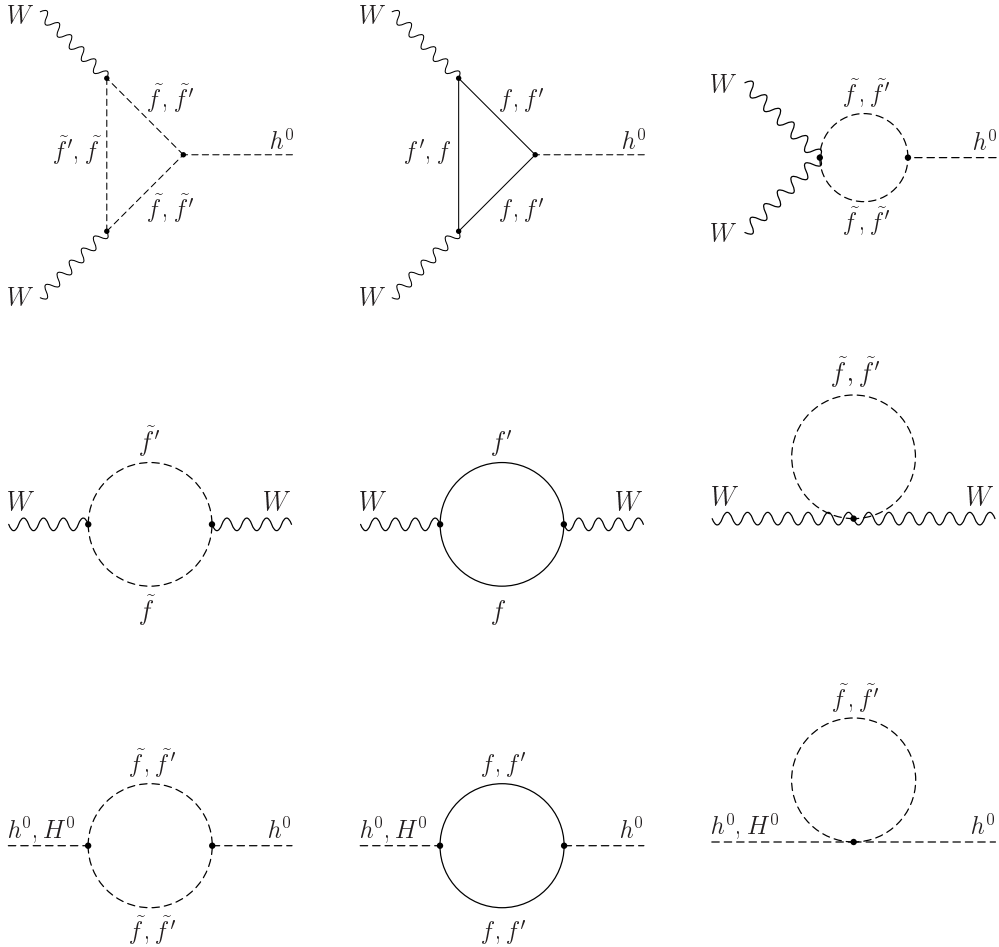}  
\caption[]{The Feynman graphs that contribute to the vertex 
correction to the form factors $F^{00}$ and $F^{21}$ and the 
wave-function correction to $F^{00}$.   
$f$ ($f'$) denotes the up (down) type fermion.} 
\label{fig1}
\end{figure}
%%%%%%%%%%%%%%%%%%%%%%%%%%%%%%%%%%%%%%%%%%%%%%%%%%%%%%%%%%%%%%%%
The one-loop part of the $WWh^0$ coupling can be
expressed in terms of all possible form factors
\beq
\hspace{-0.8cm}
   \left(\Delta g_{WW}^{h^0}\right)^{\raisebox{-1.3mm}{$\scriptstyle \mu\nu$}} =
   F^{00} g^{\mu\nu} + F^{11} k_1^\mu k_1^\nu 
 + F^{22} k_2^\mu k_2^\nu 
 + F^{12} k_1^\mu k_2^\nu 
 + F^{21} k_2^\mu k_1^\nu  
 + i\,F^\epsilon \epsilon^{\mu\nu\rho\delta} 
                         k_{1\rho} k_{2\delta} \, ,
\label{ff}
\eeq
where $k_{1,2}$ denote the four-momenta 
of the off-shell $W$-bosons. The full analytic expressions of the 
renormalized form factor $F^{00}$ and  
of $F^{21}$, 
which are based on the evaluation of the Feynman graphs given in Fig.~\ref{fig1}
with additional counter terms,
can be found in Ref.~\cite{EMS}. 

%%%%%%%%%%%%%%%%%% Fig 2 %%%%%%%%%%%%%%%%%%%%%%%%%%%%%%%%%%%%%%%%%%%
\begin{figure}[t]  
\centering\includegraphics[scale=1.]{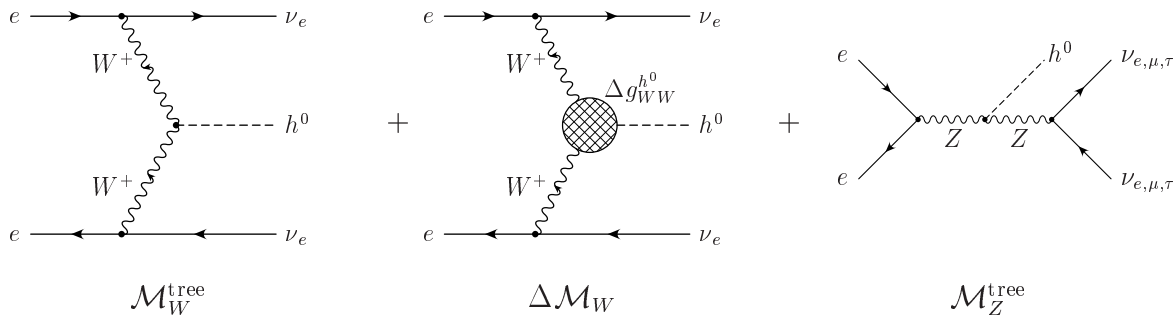} 
\caption[]{ The Feynman graphs for the process $\higpro$ including the
one-loop correction $\Delta g_{WW}^{h^0}$.
Note that for $|{\cal M}_Z^{\rm tree}|^2$ one has to sum over all
three neutrino flavors.} 
\label{fig2}
\end{figure}
%%%%%%%%%%%%%%%%%%%%%%%%%%%%%%%%%%%%%%%%%%%%%%%%%%%%%%%%%%%%%%%%
The Feynman graphs for the one-loop corrected amplitude are drawn in 
Fig.~\ref{fig2}. The tree-level part of the squared amplitude was already 
calculated in Ref.~\cite{kilian}, and the expresions for the one-loop part, 
$2\, \Re\left[\Delta {\cal M}_W  \left({\cal M}_W^{\rm tree}\right)^\dagger\right]$ 
and $2\, \Re\left[\Delta {\cal M}_W \left({\cal M}_Z^{\rm tree}\right)^\dagger\right]$, 
can be found in Ref.~\cite{eberl}. 

For the calculation of the cross section at 
tree-level, it is possible
to perform some of the phase space integrations analytically and the rest
of them numerically \cite{alta,kilian}. 
However, including  the one-loop correction
terms, it is
impossible to perform any of these integrations analytically~\cite{eberl}.

%%%\section{Results-Discussion}
\vspace{2mm}
Now let us discuss to the numerical results.
The tree-level $WWh^0$ coupling for values of $\tan\beta > 5$ as preferred
by the LEP Higgs boson searches,  mimics the SM one. 
For the calculation of the fermion/sfermion one-loop
corrections to the $WWh^0$ vertex, the contribution
of the third family of fermions/sfermions has been taken into
account. This contribution turns out to be the dominant one, in
comparison with the first two families corrections, due
to the large values of the Yukawa couplings $h_t$ and $h_b$.
The effect of the running of the coupling constants $g$ and $g'$
has been taken into account.

%%%%%%%%%%%%%%%%%%%% Fig 3 %%%%%%%%%%%%%%%%%%%%%%%%%%%%%%%%%%%%%%%
\begin{figure}[t]  
\begin{center}
\includegraphics[scale=.75]{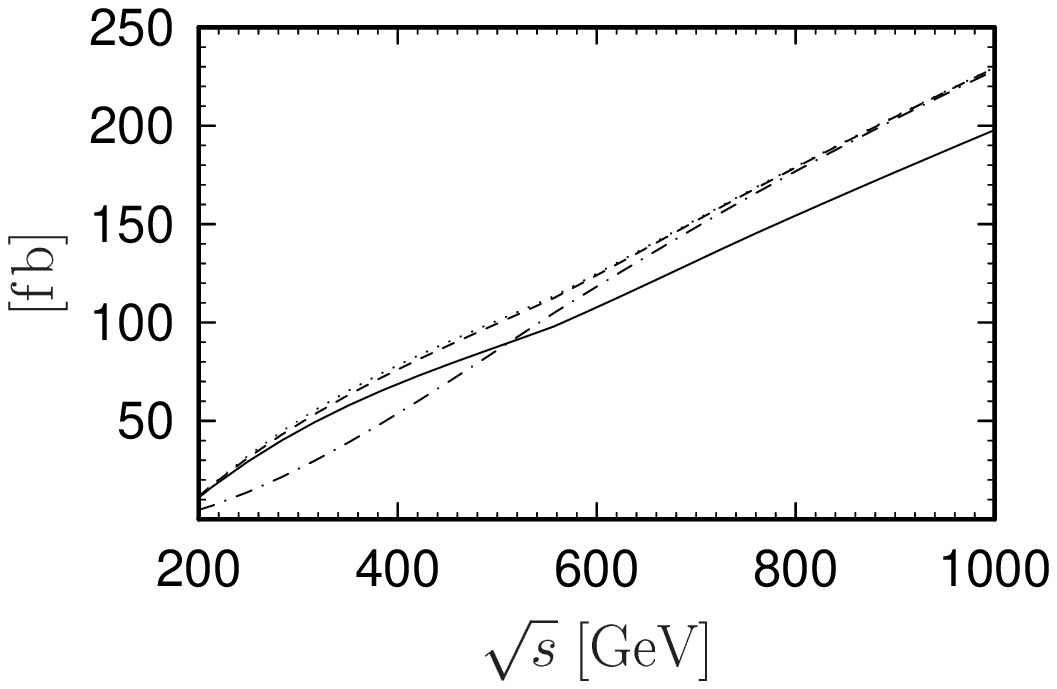}
\includegraphics[scale=.75]{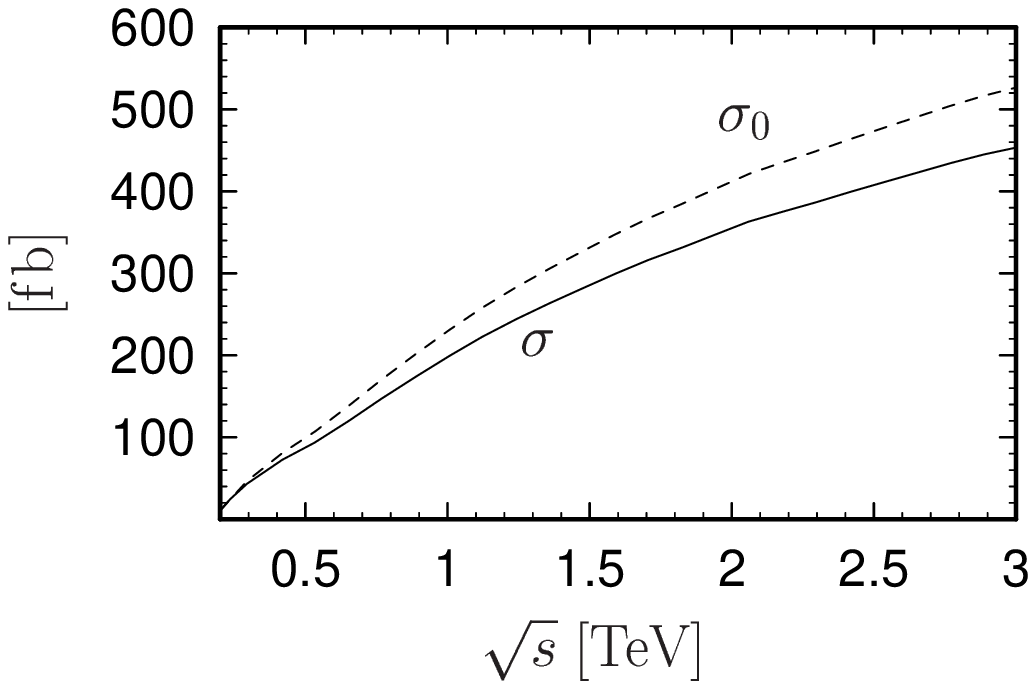} 
\end{center}
\vspace*{-0.5cm}
\caption[]{The various cross sections as functions of $\sqrt{s}$ (left).
The dotted-dashed line represents the  
tree-level cross section $\sigma_0^{WW}$, 
the dotted line 
$\sigma_0^{WW}+\sigma_0^{h-str}  $. The dashed line 
includes also the interference term $\sigma_0^{\rm interf.}  $ and
represents the total tree-level cross section.
The solid line
includes  the  one-loop correction.
The SUSY parameters are:
$\tan\beta=10$, $\mu=-100\GeV$, $A=-500\GeV$,   
$m_{\tilde{Q}}= 300\GeV$, $M_A=500\GeV$, and $M_2=400\GeV$.
For the same set of parameters also the
tree-level cross section $\sigma_0$ and the one-loop corrected
$\sigma$  are plotted for  $\sqrt{s}$ up to $3\TeV$ (right).
} 
\label{fig3}
\end{figure}
%%%%%%%%%%%%%%%%%%%%%%%%%%%%%%%%%%%%%%%%%%%%%%%%%%%%%%%%%%%%%%%%%%
For simplicity, for all plots we have used
$A_t=A_b=A_\tau=A$, 
$\{m_{\tilde{U}},m_{\tilde{D}},m_{\tilde{L}},m_{\tilde{E}}\}=$
$\{ \frac{9}{10},\frac{11}{10},1,1 \} \, m_{\tilde{Q}} $
and  $M_1=\frac{5}{3}\, M_2\, \tan^2 \theta_W$.
The choice of a common trilinear coupling and
the correlation between  the soft sfermion masses 
are inspired by unification. 

\vspace{2mm}
In Fig.~\ref{fig3} the parameters $\tan\beta=10$, $\mu=-100\GeV$,  $A=-500\GeV$, 
$ m_{\tilde{Q}}= 300\GeV$,
$M_A=500\GeV$, and $M_2=400\GeV$ are taken. Note that choosing different sets of parameters,
the basic characteristics of these plots remain indifferent.
In the left figure the various cross sections
as a function of $\sqrt{s}$ for values up to $1 \TeV$ are plotted.
The dotted-dashed line represents the contribution from
the $WW$ channel at tree-level alone, whereas the dotted line includes
the Higgsstrahlung contribution as well. The dashed line
comprises in addition the interference between the $WW$ channel and
Higgsstrahlung. One can perceive that the size of this
interference term is extremely small, and for this reason
the difference between the dotted and dashed lines is rather minute. 
For $\sqrt{s} \gtrsim 500 \GeV$ the
$WW$ fusion contribution  dominates the total cross section
for the Higgs production $\higpro$. Actually, for  
$\sqrt{s} \gtrsim 800 \GeV$  the total tree-level cross section
is due to $WW$ fusion.
In the solid line we have taken into account the
one-loop correction from the fermion/sfermion loops.
The right figure of Fig.~\ref{fig3} shows the tree-level cross section $\sigma_0$
(dashed line) and the one-loop corrected
cross section $\sigma$  (solid line) for
energies up to $3 \TeV$. Both figures show that the correction is always negative
and in the order of $10\%$.

%%%%%%%%%%%%%%%%%%% Fig 4 %%%%%%%%%%%%%%%%%%%%%%%%%%%%%%%%%%%%%
\begin{figure}[t]  
\begin{center}
\includegraphics[scale=.75]{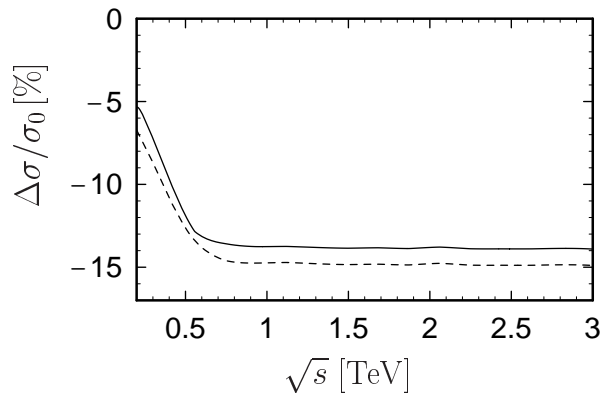} 
\end{center}
\vspace*{-0.5cm}
\caption[]{The relative correction $\Delta \sigma / \sigma_0$
as a function of $\sqrt{s}$ ($\Delta \sigma= \sigma - \sigma_0$),
where $\sigma_0$ is the tree-level and $\sigma$ the
one-loop corrected cross section.
The solid and dashed lines correspond to two different
choices of the SUSY parameters, as described in the  text.
} 
\label{fig4}
\end{figure}
%%%%%%%%%%%%%%%%%%%%%%%%%%%%%%%%%%%%%%%%%%%%%%%%%%%%%%%%%%%%%%%%%%
\vspace{2mm}
In  Fig.~\ref{fig4} shows the relative 
correction $\Delta \sigma / \sigma_0$ as a
function of $\sqrt{s}$.
The solid line corresponds to the set 
$\tan\beta=10$, $\mu=-100\GeV$, $A=-500\GeV$,  
$ m_{\tilde{Q}}= 300\GeV$, $M_A=500\GeV$, and $M_2=400\GeV$,
whereas for the
dashed line 
$\tan\beta=40$, $\mu=-300\GeV$ and $A=-100\GeV$ was taken, keeping
the rest of parmeters unchanged.
We see
that the size of the one-loop correction to the
Higgs production cross section becomes practically
constant for $\sqrt{s} > 500 \GeV$ and weighs about $-15 \%$,
almost independently of the choice  of the
SUSY parameters.
This is due to the fact that the one-loop
corrections are dominated by the fermion loops, and
therefore the total correction is not very sensitive
to the choice of the SUSY parameters.

\begin{figure}[h!]  
\begin{center}
\includegraphics[scale=0.75]{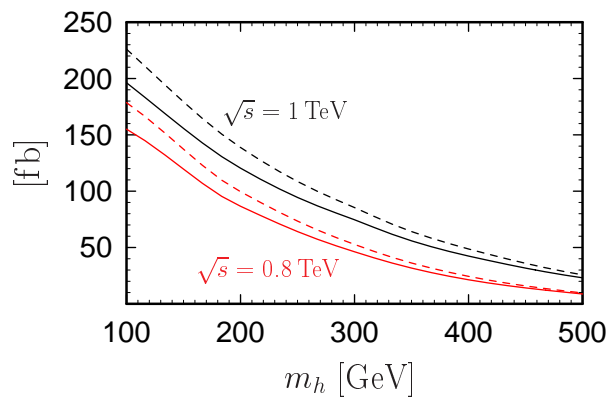} 
\end{center}
\vspace*{-0.5cm}
\caption[]{Cross sections for the  SM case, 
 for  $\sqrt{s}= 0.8 \TeV$ (red lines) 
and $1\TeV$ (black lines). The dashed lines correspond to
the tree-level cross section, whereas
the solid lines to the one-loop corrected one.
} 
\label{fig5}
\end{figure}

\vspace{2mm}
In Fig.~\ref{fig5} we have plotted the cross section 
as a function of $m_h$ for
the SM case, for  $\sqrt{s}= 0.8 \TeV$ (red lines) 
and $1\TeV$ (black lines). The dashed lines correspond to
the tree-level cross section for $\higpro$, whereas
the solid lines contains the one-loop correction stemming
from the fermion loops. In addition, the couplings have
been adjusted to the SM corresponding couplings.   
The plot exhibits the expected dependence of the cross
section on $m_h$. What must be noticed is that
especially for small Higgs boson masses $ \lesssim 200 \GeV$,
the size of the fermion loops correction becomes important
for the correct determination of the Higgs boson mass.
%%%%%%%%%%%%%%%%%%% Fig 5%%%%%%%%%%%%%%%%%%%%%%%%%%%%%%%%%%%%%%%%
%%%%%%%%%%%%%%%%%%%%%%%%%%%%%%%%%%%%%%%%%%%%%%%%%%%%%%%%%%%%%%%%%

%%%%%%%%%%%%%%%%%%%%%%%%%%%% Fig 6  %%%%%%%%%%%%%%%%%%%%%%%%%%%%
\begin{figure}[t]  
\begin{center}
\includegraphics[scale=.75]{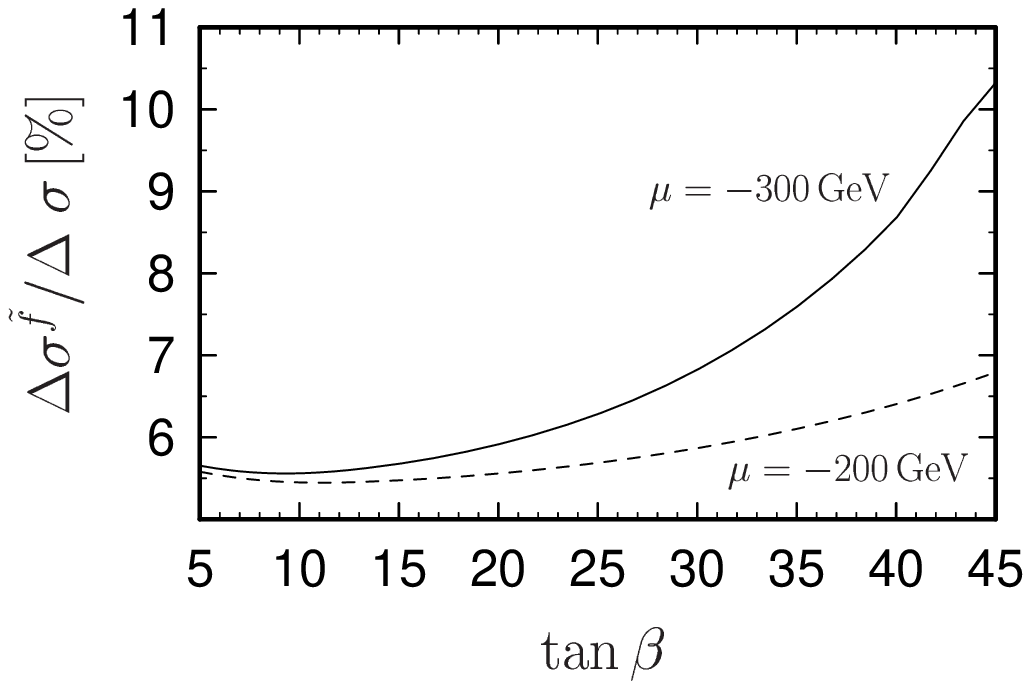}
\includegraphics[scale=.75]{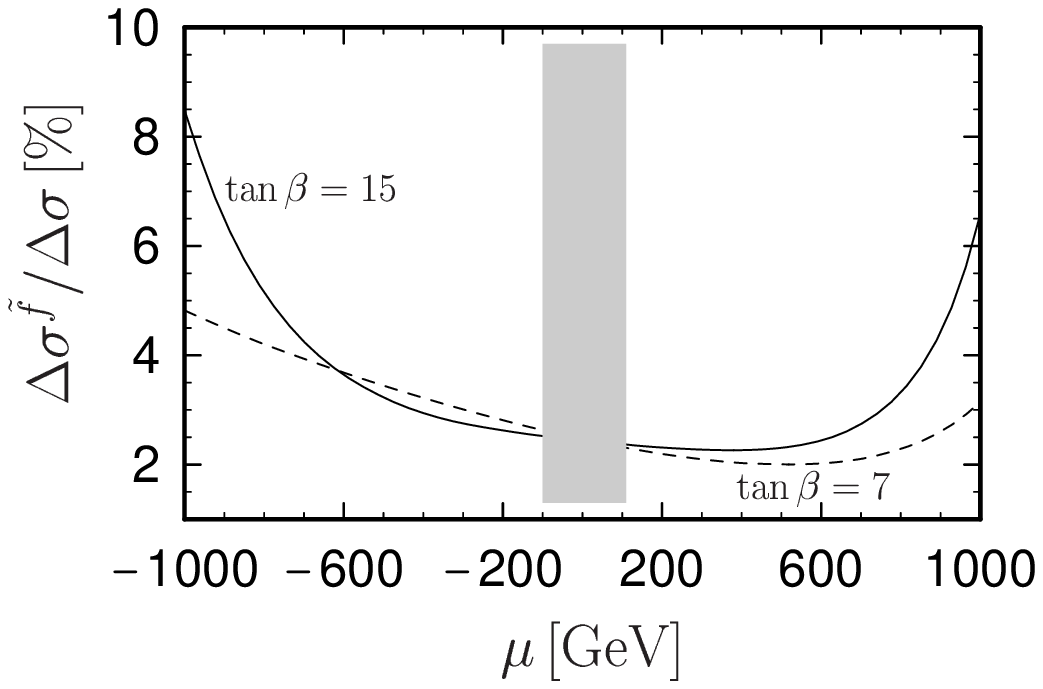} 
\end{center}
\vspace*{-0.5cm}
\caption[]{The percentage of the sfermions to the total one-loop
correction as a function of $\tan\beta$ (left) and  $\mu$ (right).
The rest of the SUSY parameters have been fixed as described in the
text. Here  $\sqrt{s}=1\TeV$.
The grey area in the right figure is excluded due
to the chargino mass bound.
}
\label{fig6}
\end{figure}
%%%%%%%%%%%%%%%%%%%%%%%%%%%%%%%%%%%%%%%%%%%%%%%%%%%%%%%%%%%%%%%

\vspace{2mm}
Finally,  Fig.~\ref{fig6} exhibits the percentage of the
sfermion loops to the total one-loop correction as a function
of $\tan\beta$ (left) and $\mu$ (right), for
two different values of $\mu$ and $\tan\beta$, respectively, as
shown in the figure. 
In the left (right) figure $A=-100\GeV$ ($A=-400\GeV$).
Further we have
$ m_{\tilde{Q}}= 300\GeV$, $M_A=500\GeV$, and $M_2=400\GeV$, and
$\sqrt{s}$ has been fixed to $1\TeV$.
The grey area in the right figure is excluded due
to the chargino mass bound.
One can see that the maximum value
of order 10\% can be achieved for large values of $\mu$ and
$\tan\beta$. There, due to the significant mixing in the stop and sbottom
sector, the  contribution of stops and 
sbottoms in the loops is enhanced.
Even for such a parameter set
the dominant correction, at least $90\%$ of the 
total correction, is due to the fermion loops.\\
An effective approximation of the one-loop corrections to the 
$WWh^0$ vertex can be found in \cite{kniehl-review}, but it does not
fully account for the whole effect. 
    
%%%\section{Conclusions}
\vspace{2mm}
In conclusion, we have calculated the fermion/sfermion loops
corrections to the single Higgs boson production $\higpro$ in
the context of the MSSM and SM. 
They are supposed to be the dominant radiative corrections.
For energies relevant to the future
linear colliders, $\sqrt{s} \gtrsim 500 \GeV$, the $WW$ fusion
channel dominates the cross section. 
In general, the  correction due to fermion/sfermion loops is negative 
and yields a correction to the cross section of the order of $-10\%$.
The bulk of this correction stems from the fermion loops, and
usually turns to be more than $90\%$ of the total correction.
For the case of  maximal mixing in the sfermion mass matrices,
the contribution of the sfermion loops is enhanced, but
nevertheless weighs less than $10\%$ of the total one-loop correction.
As the correction is dominated by fermion loops  and is rather independent
of  $\sqrt{s}$ for $\sqrt{s} > 500 \GeV$, 
we think that  it can be approximated by a 
factor correction to the \mbox{tree-level} cross section. Such an
approximation would be most useful for including initial state 
radiation (ISR) and beamstrahlung in an efficient way.\\ 

\noindent 
{\bf Acknowledgements}\\
\noindent 
V. C. S. acknowledges support by a Marie Curie Fellowship of the EU
programme IHP under contract HPMFCT-2000-00675.
The authors acknowledge support from
EU under the HPRN-CT-2000-00149 network programme and the ``Fonds zur F\"orderung der
wissenschaftlichen Forschung'' of Austria, project No. P13139-PHY.

\end{document}